# Federated Fair Trade Energy: Speculative Fabulation for a Planet with Limits

Dawn Nafus†
Department of Anthropology
Oregon State University
Corvallis, Oregon USA
nafusd@oregonstate.edu

Laura Watts†
Center for Applied Ecological Thinking
University of Copenhagen
Copenhagen, Denmark
lw@hum.ku.dk

## ABSTRACT

What happens to permacomputing when electricity grids shift to decentralised green energy, and local communities and municipalities have increased governance over this vital public service? Electricity and computational networks are more than just separate systems that plug together. Shifts to renewable energy generation in the grid are impacting computational systems, and computational demands on electrical power are impacting the electricity grid; one infrastructure limits the other. Permacomputing research tends to focus on 'off-grid' or 'behind-the-meter' energy. This sacrifices some of the social justice benefits that the public electricity grid, managed and regulated for universal service, was designed to provide. Our paper uses empirically-grounded 'speculative fabulation' to identify research opportunities in permacomputing that open up when energy systems are federated across communities. The speculative fabulation takes the form of an interview between the energy manager of a future energy island in the North Sea and a permacomputing podcaster. This allows us to conceive of computing-grid integration in tractable social and ecological terms, and introduce a notion of 'fair trade energy'.

† The authors contributed equally to this paper and are listed alphabetically.

## 1 INTRODUCTION

The limits of computation can be considered an energy problem, whether in terms of limited materials, manufacturing, maintenance, or electrical power. Computational systems do more than just use electricity; they are entangled in electricity networks. National power grids are constituted by cables with material limits and this *ipso facto* means that the computation it powers is also limited. This limit is the reason hyperscale data centres are turning to on-site fossil fuel generators: they want more energy than the grid can provide. The choices and compromises made in energy systems–choices by energy operators about what and where to generate electricity, and whom to sell power purchase agreements–shapes the computational systems that depend on them. How green, sustainable, or ethical an energy network claims to be is also a claim on the computing systems it powers, either in carbon metrics or in ecological and sociomaterial consequences.

Sustainability-oriented computer scientists have considered the implications of a green electricity grid for computing. Carbon-aware computing [1], power-proportional networking [2], and data centre-grid coordination [3] are three examples of areas that have emerged in response to a changing grid. Conversely, scholars focused on the green energy transition are turning their attention to the impact of power demands from computational systems [4]. For both, attention is often on the problem of coordinating energy supply and demand, which relies on grid management software and remote asset control. These keep electricity demand balanced with electricity generation. Without these controls, sudden swings in electricity supply or load–peaks from unexpected wind resource or from massive data centre demands–can wreak havoc on the grid network leading to blackouts [5]. In these ways, energy and computation are more than just separate systems that plug together. They can be considered as a single infrastructure that evolves [6] or a relationally-aware 'digital energetic' system [7].

If energy and computing systems are to couple in ways that operate within limits, sustainably and equitably, this requires design intention. It is a complex and emergent system. Attempting to describe the ever-changing whole, or un-black-boxing subsystems (e.g. focusing on data centres or smart grids) may not provide novel design insights. Here we choose to make different assumptions about how an energy system might operate–assumptions grounded in ethnographic realities but far from commonplace–in order to open new areas of enquiry.

Our interest is in scoping what computing systems could look like if we have a more 'desirable' energy system. For us, 'desirable' is a situated future [8], one that is located in our knowledges in the Global North and in our empirical research. We do not try and define a planet-wide desirable future for everyone. From Watts' research with energy islanders in the Scottish islands of Orkney, and their struggles for energy self-determination, we choose a desirable future that is a decentralised electricity grid network that has some local governance but remains an interconnected public service [9]. As with Orkney at present, the local electricity grid is reliant on renewable energy, principally wind, but also wave, tide, solar, and biomass. Since 2013, the islands have generated over 120% of their annual electricity needs from renewables, with local communities benefiting from selling their electricity to the grid. Our future is decentralised because renewable energy as an environmental resource is physically dispersed and often in locations like this, at the geographic edge [10].

We have created a speculative fabulation based on long-term ethnographic research to flesh out the specificities of what a 'desirable energy system' could look like. We then reflect on the implications of this system when it becomes coupled with computing. Even if our situated future does not overlap with readers' notions of a desirable climate and socially resilient future, our thought exercise nevertheless enables the LIMITS community to think more expansively about how grids could be organised differently when coupled with computing, and how computing within limits approaches might better integrate with grids, rather than be decoupled from them.

## 2 HOPE AS METHOD

We write this paper as counterpoint to the often dystopian imaginaries of the future that circulate, and instead work to establish a future for green energy and computation systems that we find liveable. In this we draw on demonstrated systems that redistribute power, in both a political and energetic sense, and approaches that address energy and data injustice such as permacomputing and solarpunk [11].

Our paper enacts hope as a method [12], where hope is not a passive wish but an empirical practice. We enact hope for digital energetic futures through a method of speculative fabulation [13]: we craft our speculative fabulation based on our combined empirical research inside the European energy and US tech industries. Watts is a science and technology studies scholar who has spent two decades collaborating with energy professionals and small communities making the green energy transition on the Orkney islands [9]. Most recently Watts co-designed the data governance of Responsive Flexibility (ReFLEX) Orkney, the demonstration of an Integrated Energy System that balanced energy across heat, transport, and power in the islands [14]. Nafus is an anthropologist who has conducted research and development at a US chip manufacturer also for two decades. For this work she extrapolates from applied ethnographic research in data centre energy measurement [15], and carbon-aware scheduling [16], orchestration [1], and networking [17].[1] Both our speculative fabulation and the conclusions we draw are grounded in this body of empirical research.

To be clear, our speculation is not fiction but a thought experiment that acknowledges how knowledge-making requires partial creative practice [13]. It is when "the study and crafting of fiction and fact happen explicitly, instead of covertly, in the same room" [18]. This is resonant with speculative design [19], which explores how futures can be otherwise through speculation grounded in design practice. Rather than write a paper that gestures towards a future, we want to take you there. We want to make it practical, so you can grab hold, and not just contemplate it from afar.

We are located in the Global North where most people live with the privileged expectation that both power and digital networks are 'always on'. Here, many national and regional electricity grids are shifting from centralised fossil fuel power stations to decentralised renewable energy, which is distributed around the energy network edge (where the wind is, for example), often in remote locations and communities. There is also a growing regulatory and cultural commitment to supporting local energy governance (such as the European Union's Citizen Energy Communities), raising questions about how computing could be redesigned if local governance were the default assumption of how electricity was delivered, and not just one of several national initiatives.

## 3 RELOCATING THE ENERGY-COMPUTE LIMIT

Changing the default assumption that energy is a centralised system, towards assuming one that requires local governance, demands consideration of scale. At what scale should we speculate: what are the limits of the ‘local’ in an energy-compute network? This is not straightforward because energy and computational networks do not just scale up or down together.

[1] These reflections are based exclusively on information that has been published or is otherwise widely known within the technology industry, and in no way reflects any potentially commercially sensitive information.

To have a notion of computing within limits is to have a notion of a boundary: a notion of what counts as 'inside' and 'outside'. In this paper we use the ‘regional' to define a place-based boundary that is not necessarily administrative, like a municipality or state, but might overlap with those borders. When we talk about 'community' we mean it as a permeable boundary rather than as a defined group. Neither 'regional' or 'community' are immutable wholes, yet they matter socially. They also point to a disjuncture between current permacomputing boundaries and how electricity grids work.

Coarsely speaking, notions of energetic limits in permacomputing tend to be attached to one of two scales. One is planetary scale, where notions of 'limits' echoes and extends the classic *Limits to Growth* thesis, sometimes drawing on the notion of planetary boundaries from climate science [20] or doughnut economics [21]. These works raise vital questions about how much, of what kind, of computing is needed [22] [23] [24], similar to the way United Nations negotiations focused on overall fossil fuel drawdowns.

The other notion of ‘limits’ attends to the hyperlocal, or site-specific, as a way of instantiating what 'within limits' could be. In electricity terms, these are *behind-the-meter* places, meaning the electricity grid is not a factor in generating or managing power; its work stops at the meter. For example, the Windternet project [25] describes itself as ‘grid liberated’. The project pushes how far degrowth computing can go using mycelial wind turbines to power a small server, which hosts agricultural information for the local farming community. Another example is the Solar Protocol, which demonstrates that a website can be supported using exclusively intermittent, but coordinated, on-site solar electricity [26].

These two examples do engage with notions of community. Communities participate as users of locally relevant information, or as a distributed community of people who maintain small solar and compute resources. Communities as local entities that manage and transmit electricity off-site, however, are largely absent in permacomputing. One exception to the pattern is Soden et al [27], which is intriguingly ambivalent about electricity grids. Its work takes place across two sites. One site is in Scotland, where community energy is supported by and facilitated by governmental entities. Here, the text points to inequalities that can emerge in 'off-grid' futures, and the involvement of government implies that some coordination with grid operators is likely. Another site, a speculative fiction set in Toronto, sees community control over energy happening entirely behind-the-meter, where people swap batteries filled from rooftop solar as best they can because the regional grid has abandoned them.

Overall, though, it appears that a community-controlled electricity grid is hard to imagine within permacomputing. There are practical reasons for this. The tight regulation of electricity grids makes it impossible to share one's wind or solar energy with one's neighbour down the road, whether for computing or other use. As an individual customer you are located behind-the-meter and your relationship with your energy operator is one of them supplying your home or, in some cases, you selling electricity from your home to your operator. The prospect of a collective and local electricity system is therefore unlikely to occur to many. When we want to reduce the scale of something as massive as 'planetary boundaries’ researchers are understandably drawn towards the individual household site as the appropriate spatial boundary.

Well-run electricity grids provide an important baseline public service, however, which means they matter for energy justice. So we need to reconsider the limits of energy-compute systems in ways that allow us to engage with grids at the scale they operate at. For electricity grid management, there are two scales that matter: the wide-area or national (transmission system) and the regional (distribution system). Broadly speaking, the national or transmission grid operator manages the high-voltage lines that transport electricity long distance (known as wide-area interconnections in North America), and the regional grid operator manages the lower-voltage lines that carry power to individual households and businesses. Everything in the electricity business is about the regional 'limit' of the network and how it interconnects with the network ‘next-door'–such as a neighbouring national or interconnection grid. For a visual impression of this regionalisation look at the live map hosted by ElectricityMaps.[2]

The scale of both of these networks is regional and communal. Neither is individual–the electricity industry's relationship is not with individuals but with the electricity meter, and the collective that lives behind that meter, whether a household or multi-building campus. Neither is planetary. National electricity networks were established as public services to provide equitable and efficient access to power, and then these services were privatised in the late twentieth century. National regulation remains, and costs in one place are still socialised across a whole region. Although electricity interconnector cables allow electricity to flow across regional borders–even whole countries, there are no interconnectors across continents, or around the globe as with undersea internet cables.

[2] Hosted at https://app.electricitymaps.com/map/live/fifteen_minutes.

This history means that communities have not been in full control of their grids, even if in some places they can exercise important amounts of agency. Some countries, such as Norway, do have smaller-scale regional operators but the lack of local grid control is a sore spot in many places where blackouts and brownouts do occur [8]. In the shift to decentralised renewable energy, grids have retained their public service obligation but are also exploring diverse forms of regional and community-scale governance, from Virtual Power Plants to Community Choice Aggregation [28] [29]. Interconnected microgrids have been proposed as a vital component of climate resilience [30]. These small-scale systems, which aggregate assets together, have arisen as one response to the increased complexity of decentralised renewable energy. In the past, grid managers only had a handful of centralised power stations to worry about; now, in a green powered grid, there are thousands of wind turbines, solar panels, and hydrogen fuel cells, which all have to be managed together.

Our speculative future works at this community scale because community-informed governance increases the pace of the green energy transition. For example, Watts has shown that local-owned wind turbines and the high levels of wind in the landscape are part of strong relationship with energy as a whole in Orkney, and give communities a sense of ownership of 'their' local electricity. People want to participate in governing 'their' resource [9]. More generally, community energy governance increases social adoption of renewable energy and negates so-called NIMBY-ism [31].

For our speculative fabulation, we use Watts' empirical research on North Sea energy islands, Orkney, as the basis for our region. But instead of rehearsing that present, we speculate on an island in the North Sea that is one possible future.

That future is already being built. The European energy industry is designing and building artificial energy islands in and around the North Sea to manage and distribute its gigawatts of offshore wind power [32], the most notable being Princess Elisabeth Island and the Bornholm Energy Island. These are energy island communities yet to come. These artificial energy islands are our inspiration for ‘Atlantis'–the site of our speculative fabulation.

So, we invite you to listen and read an interview between the energy manager of one energy island in the North Sea (Charlie), and the interviewer from the permacomputing podcast, Permabits (Dawn).

## 4 PERMABITS PODCAST

### Interview with the author of the manifesto for federated fair trade energy[3]

*Dawn*: Welcome to Permabits, a podcast for perma-computing. You might have heard about the Atlantis Manifesto, which sets out some interesting principles for fair and federated energy systems. Green power is one of the foundations of permacomputing, so today we’re interviewing Charlie Bevan who wrote that manifesto. She lives on an island in the North Sea at the heart of a new energy system, an island which is–unexpectedly for an island in the North Sea–called Atlantis.

So, Charlie, for those of us who don't know what Atlantis is, tell us a little bit about your island, and how it inspired the manifesto.

*Charlie:* Thank you for the invitation, Dawn. Atlantis is an artificial energy island in the centre of the North Sea. We were built by the European Union to manage the gigawatts of renewable energy being generated from its offshore wind turbines. There are thousands of wind farms in the North Sea but the wind is intermittent–sometimes it blows, sometimes it doesn’t–and people want electricity at completely different times. Demand does not align with generation. So we were designed to switch North Sea wind energy to a national grid where there is demand–from Norway to the Netherlands–or to store wind energy when nobody wants it. We manage North Sea electricity so that when you put your kettle or computer on in this part of the world, you will get wind energy even if the wind isn't blowing.

My main job is to help balance the European grid by shifting energy in time and space. So Atlantis has grid batteries, hydrogen electrolysers, and even data centres, all of which transform energy from one form into another: electrical energy into hydrogen or data, for example.

But I have another job because we are an island community, too, with our own energy needs. Atlantis has its own electricity network to keep our home and office lights on, to keep our geodesic greenhouses irrigated, and our lighthouse beacon flashing out over the fog. We have electric bikes and home batteries and servers and hot water tanks, which all need power. I think it's important for your listeners to know this: my neighbours, they come first. The island community has to come first because we are precarious. We are a hundred or so people living out here on a lump of concrete and sand in the middle of the North Sea. Our kids are here. This is our home. We are going to stay. But it’s not easy.

[3] The Atlantis Manifesto can be read at https://sand14.com/fair-trade-energy/

So there are two energy systems on Atlantis. We make different choices about how we distribute and manage our island energy amongst ourselves, compared to the choices we make to manage the gigawatts of hyperscale wind energy that flow through us. We don't govern Europe's energy. But we do govern our own energy. So we can–and we do–give our energy away for free if we choose. We made a set of principles to help us make decisions about what's best for us. Those principles were the first step towards the Atlantis Manifesto.

*Dawn:* The Atlantis Manifesto talks about trading energy between places. You talked just now about the difference between the European energy system and the island's energy system. These all have different limits, different system boundaries.

We talk a lot about limits in our permacomputing community. One limit that we've experimented with is what you call behind-the-meter, where the limit of the energy network is the electricity meter. We might put up solar panels and servers that are linked together around the world to run a solar-powered website, for example. That's about exchanging data. But what we don't do is think about exchanging energy, which is what the Atlantis Manifesto seems to be about. Can you explain that to me, and our listeners, where the limits are in your system? Is it an exchange between different local energy systems, like behind-the-meter? Or is it like exchanging energy with a single central company, or one per country? How far and wide do you go?

*Charlie:* Well, we are a grid, so we don't get involved in anything behind your electricity meter, like the projects you mentioned. Electricity grids also stop at the end of their grid cables, and we don't have electricity cables sending power around the planet because it's physically not sensible–you lose too much en route. There are no transatlantic electricity cables like there are internet cables. We can just about manage cables across the North Sea. Electricity grids are regions. They are always regional-scale, whether that be a country or an island of a hundred people. I'm an energy nerd so that's the scale I always think in: regions, and sometimes those regions become communities.

Here on Atlantis the island is the limit. Islands are handy like that: it's always obvious where your limit is.

I'm going to explain our island grid in simple terms. So apologies to the energy engineers out there. We run our own island electricity grid. All our clever kit and hot water tanks and kettles: we keep them balanced so we can present ourselves as one simple energy asset to the European grid. We say, "don't worry, big electricity grid, we've got this". We have aggregated all our energy equipment into what's called an Integrated Energy System. That means we balance our energy across, not just electrical power, but also transport and heating. The European grid treats us like one big house. The entire island is behind-the-meter, so to speak.

That means that we get to make the rules for ourselves inside our system–inside our house. We can do what we want. That's how we can decide to give energy away for free to islanders or prioritise those with medical equipment.

Outside our system we plug-in to the European electricity grid and its rules, so it's not like everything goes. But you'd be surprised how creative your energy governance can get on a small island. Fuel poverty? We made sure that never happened.

For me, the limit is the boundary of what you can govern. I don't mean me, personally, of course. We have an energy co-operative here on Atlantis that makes the decisions. But energy communities can choose whatever governance works for them. And listeners should know this is not new. There's been energy co-operatives for almost a hundred years. Then, in 2018, the EU introduced special company structures just to manage local energy systems. The tools for fair trade energy have been around for ages.

*Dawn:* It sounds like you've drawn the limit in an intelligent way: it's not an individual house and it's not the entire world. The limit is an island, which is then in a relationship with other communities that are, in a way, also like islands.

I love the idea of my community governing their own energy system, but I don't think we have an energy nerd here. In the permacomputing world, we need certain kinds of nerd, but computing nerds and energy nerds speak a different language. Frankly, it's incredible to think about the technical skill you need to run a community-scale energy system. It's just so much more complicated than a 'me with my solar panel' setup.

*Charlie:* Skills are actually foundational to us. We're pragmatic and practical. I think that comes from living on a remote island: you always have to fix things yourself, probably with a coat-hanger. Fair trade energy isn't about trading numbers in a spreadsheet. For us, fair trade is a lot about trading skills.

Resilience is a word that's everywhere, but resilience isn't just a word, it's something you have to do. In our climate-changed times resilience is something we all need to grow. Training, learning new skills, learning how to manage your energy system, that's growing your resilience when you live in a place that's precarious–and that's many more places than you hear about in the news.

So, you are right, you need energy skills. And, in fact, your situation is how our federation of energy communities started.

We were contacted by a friend of mine who lived in what had been a fishing community on the North Sea coast. They had a giant substation for one of the offshore wind farms, but that electricity was doing nothing for them–it just *wooshed* past, so to speak. Half the people there were in fuel poverty; burning their floorboards to stay warm.

They wanted to do what we're doing: run their own energy network for their benefit, to fix all that. They had the electricity from the offshore wind farms passing through. They had some big thermal heaters in their schools, some solar panels, the boats had batteries. From a 'limits' perspective they were at the same scale as us. They were a small community plugged in to a giant green electricity grid. And we thought, well, if they can find someone with an interest then they can come and apprentice with me.

So they found this great boat engineer, who came and stayed on the island for a while. And she kept asking all these awkward questions. She made us realise we had to do two things. First we had to 'comment our code' (I think that's what you call it) and explain what we were doing and why. That was when we created our ten principles for fair energy governance. Then we needed to figure out how to trade with them.

They didn't want us to train their energy manager for free but they couldn't pay us in money. They could pay us in green energy, however, from inside their new integrated energy system. But trading energy is easier and harder than you think. It's easier because green electricity isn't really a material thing like bananas or coffee beans. Trading green electricity on the market for renewable or low carbon energy is actually pretty distinct from the interconnected electricity cables. It's more of a spreadsheet problem than it is an electrical problem, which is good. But energy markets are also horribly, horribly complicated, which is bad. We wanted to keep the exchange between us, between two communities who cared about similar things, and nothing like the corporates trading their green gigawatts. So we decided to keep it simple and made our own energy community exchange.

Rather than being limited to buying and selling green electricity in kilowatthours, which is what happens on the energy market, in our federation you can decide how and what to trade. You can decide that training is worth energy. So we can trade energy or–and this happened–you can pay us in upskilling our gardeners to grow lemons in the middle of the North Sea. This is our approach to trading fairly.

Trust me, having fresh lemons out here is really transformative. Food is energy: growing, nutrients, transportation. I know listeners are thinking about computing and software but, for us, that's inseparable from energy. You have to think of energy systems as a kind of ecology, which I guess is related to permacomputing: trading energy for lemons.

*Dawn:* This almost gets close to a barter system. As you say, it's what enables you to trade energy for skills, or energy for lemons. It's entirely different from renewable energy certificates that are geared to make the ESG report look good, which our listeners from the tech industry might be familiar with. Can your fair trade energy be used like renewable energy certificates are used? Can I trade with your fair trade energy exchange?

*Charlie:* To trade with us you need to join our federated energy community, which you and your community would be very welcome to do. That would mean upholding our principles but you retain governance and decide how to enact those principles. We know that what is fair in one place is not necessarily fair in another. We know that, in practice, you have to make compromises, but they are your compromises. Do you decide to have a local owned wind turbine or peatland, for example. We don't have peat on Atlantis but we have to make other compromises: our desalinator uses energy to make potable water but also salt, which isn't great for our sea life.

We have fair trade energy tokens that we exchange, which could be used, in theory, in an ESG report. I guess you could say we barter them: fair trade energy tokens for lemon growing.

*Dawn:* Can you explain how the tokens work?

*Charlie:* So, there have been 'book and claim' energy tokens for years in the energy industry. In Europe they are Guarantee of Origin certificates. In the U.S. you have Renewable Energy Certificates. Basically, companies that make renewable energy book them to a database and then companies that want to buy green energy claim them from the database. This way you can trade green energy as 'pieces of paper' all over the world. Those certificates end up in an ESG report.

But these origin certificates don't have any way of assessing how equitable or fair, for people or the environment, the renewable energy is: an offshore wind farm run by a multinational makes just the same kind of certificates as a local-owned wind turbine in Scotland. Both have made social and environmental choices but you know nothing about them. So buying green energy certificates was not enough for us. We want to know something about the places and people involved in generating and managing the electricity. That's what fair trade energy is. It's more than a certificate. It's a kind of underground community where we can trade knowing we're all committed to doing our best for where we live.

That's what makes it different from 'fair trade' as a label that you see on food. We're a set of principles that communities uphold, however they can. We're not a standard to be audited. We have relationships of trust. We know people. We help each other. Yes, we're fallible. But everything always is. What matters is growing resilience not control. We're a federation not an empire.

So, in theory 'fair trade energy' could be in an ESG report. But because we have principles, not a standard, I am not sure we'd pass a corporate audit.

*Dawn:* It sounds like old-fashioned economic anthropology, like from the days when shells and yams were exchanged around the Pacific, and each trade meant something very specific to the trading partners, and tied them together in the process. Maybe I'm making the leap a little too far, but this seems ultimately economic, just not the way much of the world usually understands that word.

I want to circle back to computation. Help me understand the computing that's going on in your energy system. You mentioned a data centre?

*Charlie*: Every community in our federation uses software to manage their energy system. The software controls all the energy assets, from a humble home battery to a tide turbine–even power to a data centre. The software switches everything on and off to balance the community's energy system and make sure it maintains its commitments to the big national grids. In order to do that, you have to balance the demand, when everyone takes a morning shower, with what's available in your energy 'kitty' so to speak.

The energy management software is where the rubber hits the road. It's where the governance decisions made by the community about who, what, and when things get power are done in practice: because when there is not enough energy, someone or something has to get switched off.

But that software that has to live on some kind of server, which requires power. So, there's an interesting dilemma. The foundation of our energy system is a data system. But the foundation of every data system is an energy system. On Atlantis we have a data centre where we host our software but we don't have the technical skill to run it–an outside company does. So part of the reason I'm excited to talk to you today is because I'm thinking about that lack. Nobody in our federation has the skill to run a community data centre. Maybe your listeners can help?

*Dawn:* Perhaps our listeners have some ideas, but let me play with this for a moment. Say I live in one of your federated energy communities and I have some cool software that solves a bit of this digital infrastructure problem you have. Or maybe I want to build some specialist weather software. Could you, maybe, pay my energy bill in exchange for developing code?

*Charlie:* Absolutely. We would work through the energy community that you pay your bill to. So we might agree to give them fair trade energy in exchange for, in essence, your code. And internally, within your energy community, you agree with them to get a few months of free energy.

*Dawn:* Okay, but let me go further. What if there were a whole bunch of coders, across different federated communities. Could they form a development team and do clever things, like run your data centre in a new way? Could you do a larger project at this federated level?

*Charlie:* For sure. I think this is one of the most exciting things about having a federated energy community. It opens up the imaginary space of how you can collaborate together. All you need is something to share with someone else. It's not limited by cash, which is important: a lot of our federated communities don't have a lot of money. But they have a whole lot of valuable expertise.

*Dawn:* Let's go back to your dilemma, which is really at the heart of permacomputing: software requires power, but managing power requires software. We've talked so far about limits that are community scale. But is that a hard limit? Would you work with a larger company or one that was otherwise just focused on making money, no matter what? What would you want to see from them for that to even be feasible?

*Charlie:* That's a really important question. I'm standing on an artificial island, on a hill that was built by a multinational. I am looking out at a heap of rusting green container boxes that hold hydrogen fuel cells, which are huge and complicated bits of kit. We know how to maintain them. But we sure as heck didn't build them.

As I said before, we're pragmatic, we're about the possible–what can we do within our limit. We're not trying to knit our own data centre, here.

But when our data centre does go down, do you think the operator sends out an engineer in a van? We're in the middle of the North Sea. Our harbour shuts when we get a good blow. Resilience means knowing how it works, running it ourselves. It means thinking long term, skilling-up our kids, making sure we'll be here for generations.

So I am asking, what does a community data centre look like? Could we talk about community infrastructure as a service? Your listeners know how to run a solar-powered Web, which sounds a lot like a federated community platform to me. How do we change the limit from one solar panel and a laptop, to a solar farm run by a local farmer and a server running software that we need? How do we keep that going for decades?

This is what we're thinking about as an island. This is why we wrote the Atlantis Manifesto.

*Dawn:* We'll put a link to the manifesto and other references in the show notes.

So that's the challenge, Permabits folks. What can we build as part of an energy federation? How would that look different from what we have today?

I've been talking with Charlie Bevan from Atlantis, author of the Atlantis Manifesto. Thank you, Charlie. And to everyone listening: see you next time.

## 5 DISCUSSION

In this speculative interview, Charlie asks the permacomputing community how to build, operate, and maintain a community data centre. This is the kind of computing resource she is likely to see, humming along on her island. However, we should not assume hyperscale 'AI factories' every time we hear the phrase 'data centre'. Instead, we can take up her call by asking how much computing makes sense within a federated energy community, given competing needs. We might then consider the range of compute resources that could be federated in ways that echo, and make use of, this federation of communities.

Federating compute resources is not new to computing. Even AI training can be federated [33]. However, if we conceive of federation with social and geographic boundaries then the focus becomes using computing to facilitate *thriving* within social and energetic limits, without resorting to exclusively behind-the-meter strategies.

Our speculation shows how federation quickly raises matters beyond computation and energy. Federating means setting new terms of exchange and interconnection. It necessarily entails decisions about who to involve oneself with, and what those relations involve economically, ecologically, materially, and culturally. The bounds of community are deliberately not sealed off. Charlie does not naively expand the scale of what she does, but questions what should be exchanged and with whom. She comes with something to trade: local green electricity. Her energetic abundance is grounded in ethnographic reality. Orkney generates more electricity than it needs.

In this speculation, not all exchanges are the same. They depend on the relationships involved. Exchanges across federated communities and on the international energy market are quite different. Sometimes 'paid' electricity bills are exchanged for code or lemons. Sometimes there is a longer-term investment in skills for a less direct return. At other times, Charlie sells electricity on the wholesale market in the same way as any other commercial entity.

We introduced a notion of 'fair trade energy' in full awareness of the bitter contestation over the dilution of fair trade standards, where corporate co-option takes the form of proliferating weak and unverifiable labels [34]. However problematic the term, it does provide a rudimentary vocabulary for the rich variation in exchanges that can and do tie people together–a vocabulary that is otherwise impoverished by capitalist domination. Some readers might resonate with other concepts like doughnut economics [21], or with noticing the gaps and patchiness of capitalism [35].

If you are an economic anthropologist, you might recognise the delayed reciprocity, perhaps even see glimmers of the non-monetary or quasi-monetary exchange systems that fascinated mid-century anthropology [36]. Historically, common goods of shared necessity like grain, yams, or cows easily served money-like functions. The limitations of barter are far less serious than orthodox economics makes out [37]. In 2026, there is no immutable reason why electricity cannot serve as informal currency. Regardless of economic specifics, which could take different forms depending on the context, the speculation demonstrates just how impossible it is to contemplate doing compute-energy infrastructures differently without also contemplating new forms of livelihood.

For computing, our speculation suggests areas of investigation that are not exactly new, but have not been typically thought through with a community limits lens. This is worthwhile because there really are energy communities, microgrids, and integrated energy systems out in the world that are already entangled in computing infrastructure. Even when these are absent, thinking at federated community scale helps us slow down default assumptions about what infrastructure is possible. Below we list four broad areas of enquiry that our speculation appears to suggest.

### 5.1 Federating hardware resources within tractable limits

We propose thinking, not with hyperscale data centres, but with heterogenous smaller resources: from phones, to under-the-desk servers, to edge data centres in telco towers or stand-alone facilities. The question is, then, what makes sense to physically place where, for what kinds of community. A federation might strategically connect wind-rich and solar-rich places so that the computing might be orchestrated according to their different rhythms. A federation with a concentration of digital animators might want to exchange different types of computing and energy resources compared to a farming community.

Questions of ownership can be considered anew. In our speculation, the island's data centre is leased and run by an external operator, but this is unsettling because it means limited local control. It does not serve islanders' ends beyond absorbing energy at the right time, and hosting relatively small grid balancing software. However, fleets of distributed computing capacity could be managed as a shared resource even if individually owned, just as Charlie manages assets she does not own towards particular ends.

Simplicity of repair is a common permacomputing goal. In this scenario it takes on new urgency across a wider range of hardware–try repairing a telco server on a remote island! Hardware and software optimised for dynamic electricity availability would need to be a priority, as the electricity available for computing cannot be assumed to be constant in an era of climate change. Some hardware are better than others at incrementally powering down when needed [2].

The politics of land use for compute resources would no doubt change in this scenario. Communities around the world define what 'fair' governance of land looks like differently. One chief complaint of communities contesting data centre permitting is the presence of backroom deals.[4] If a fleet of smaller, more distributed computing resources served community interests, as defined by that community, and had an energy use upper bound that was also determined by that community, siting discussions for larger, non-hyperscale assets might be very different.

### 5.2 Prioritising tasks within dynamic energetic limits

It is not enough to calibrate the total computing resources available to a finite upper bound. It is also important to be able to determine what runs when and where, and this requires both social governance and the technical means of enacting that governance. Just as Charlie prioritised electricity for keeping islanders warm over earning money on the wholesale market, a digital prioritisation system is worth considering if keeping grids as close to 'always on' as possible remains a desired goal.

Fierce debates rage about the thin veneer of 'need' for massive computing capacity expansion, under the guise of invented technological arms races. These reflect a very real inability to create social agreement about how much of what computing tasks are a social priority. Given that compute is limited by its power, better rules of the road are sorely needed, even if there will never be consensus. In a crisis, like a blackout due to climate change, there is currently no good way to preserve humane social priorities for compute–for example, dialling back the harmful or useless bots scraping the internet for AI training data.

Computing is, of course, no stranger to queuing systems and schedulers, but ways of determining priority reflect the assumption that the power will alway stay on, regardless of what task is run. For example, carbon-aware scheduling and placement techniques were designed to solve the problem of intermittent renewable energy generation, but the computing tasks themselves are treated as blackboxed and unquestionable. In contrast, networking protocols are not entirely black-boxed. Packet priority is based on engineering requirements for different types of tasks. Video call packets are prioritised over, say, a document upload, regardless of the contents or purpose. This is an improvement, but a video call to a dying relative is still indistinguishable from a video meeting that could have been an email.

To have a working federated computing system that operates within energetic limits means building the capacity to agree and operationalise definitions of priority. According to Nafus's conversations with data centre operators, the primary way to do that on the commercial market today is through price and service level agreements (SLAs). These practitioners made clear that should the need to dial back electricity arise, SLAs would dictate that capacity be preserved for the highest paying customers. Breaking the SLA would involve monetary compensation. For communities, however, accountability is more serious than writing a cheque. Cash is not king. Relationships and trust are the currency. Accountability within federated communities means negotiating and explaining to your neighbours why an AI system was allowed to continue internet scraping when medical records could not be accessed.

There are examples of prioritisation systems that do take into account equitable access. At some universities, the more data centre resources a researcher uses, the more likely they might get throttled if there is not enough capacity to go round. Here, fairness is a function of how much computing you get, without making claims about the worth of the task. It could also be interesting to combine new approaches to task prioritisation with intent driven orchestration (IDO) techniques [39] where data centre users are asked to explicitly prioritise between speed or energy use reduction, and the IDO software 'right-sizes' how the task is executed accordingly.

### 5.3 Reconfiguring meso-level digital infrastructure

Lists regularly circulate about how to leave ‘big tech’ that suggest switching to software and services that have not yet been ‘enshittified’ [40] or have an explicitly pro-social remit [41] [42]. From Murena, a de-Googled phone operating system, to Proton, an encrypted European service for email and calendar, to longstanding entities like Wikipedia and DuckDuckGo, these entities provide a credible threat to so-called 'big tech' monopolies and domination. They also tend to sit at the top of 'the stack', as end-user facing applications.

[4]For example, a recent legislative proposal to allow more data centres in Oregon, where one of the authors lives, was widely condemned for evading public processes. See [37].

One question our speculation raises is how other layers of software would need to be different if a federation of energy communities were the reference point for electricity supply. We can explore this by contrasting with developer-facing software that cloud providers offer to entice developers to use their cloud services. There, the design emphasis is on extensibility and scalability, drawing on images of standardised shipping containers [43]. Much systems software carries open source licenses for interoperability purposes, but is designed by and for large corporations. Federated energy communities allows us to think about software that facilitates cooperation and exchange with other places, but does not optimise for corporate expansion in similar ways. This might also place greater emphasis on including people who are new to the trade, as evidenced by Charlie's plea for skills exchange.

The data centre orchestration software, discussed above, would be one important example, but there are others. Content delivery networks (CDNs), which store larger files (typically for streaming video), are physically closer to the point of delivery so they can avoid congestion on internet backhaul. They are typically run by commercial companies, telecoms companies, or by content providers themselves such as the BBC and Netflix. With electricity a substantial operational cost of CDNs, communities able to exert greater control over their electricity could be well positioned to team up with video content developers to provide a competitive infrastructure or service.

### 5.4 Questioning how livelihoods are made

In Charlie's interview, trade flows through people who become connected in the process of installing community-driven electricity generation, or exchanging it. They become meaningfully involved in each others' lives, not through vague commitments to notions of care, but because they must. While artists, open-source software developers, and lemon growers need more than electricity to live on, having your electricity bill paid for is not a trivial economic contribution. While speculative, the suggestion invites us to think more creatively about how socially beneficial digital work can be economically supported.

At present, very few people's livelihoods are directly tied to permacomputing. For every paid worker in a repair cafe or software engineer working for the Wikimedia Foundation, there are many more hobbyists and experimenters. Corporations take advantage of free-as-in-beer labour of open source volunteers [44] and now build AI empires on the backs of the creative labour embodied in books, images, and videos. Following Sholtz [45], Sandoval [47] contrasts open participation with platform cooperativism, where the latter work for the benefit of a defined group, and directly address the issue of precarious labour.

Permacomputing as a livelihood is rarely an area of explicit research consideration ([47] is an exception). Our speculation suggests that it could take cues from platform cooperativism, and apply them down the infrastructural stack. One example is coopcloud.tech, an infrastructure software stack built by a tech co-op. Fair trade energy might beget fair trade compute capacity. This might mean designing for those with limited computing skills while thinking more creatively about exchange. Echoing Soden et al [27], students and researchers would need to reconsider what engaging with communities looks like.

## 6 CONCLUSION

Drawing on empirical research, we created a speculative island where energy and computing are entangled, and we explored its implications for permacomputing.

We propose permacomputing expand its attention to the community and regional scale, with groups who can govern their electricity resource, forms of which exist today. If permacomputing can imagine energy management as controlled by communities, not just quasi-capitalist energy companies, we have a starting point for a fairly-determined energy limit. We show how this limit leads to designing exchange systems, including exchanging skills. Building resilience is the essence of fair trade energy.

We suggest four computational areas worth considering: federating smaller-scale hardware resources; task prioritisation that takes into account both energy limits and community-determined priorities; providing competition to mainstream digital infrastructure software; and a deeper questioning of software and hardware development as a livelihood. Our four areas implicate large subfields within computer science, and we invite others to redirect or build them out with greater specificity than we are able to offer.

Finally, we invite researchers to respond to the challenge in our speculative podcast: what does a community-based set of compute resources look like, one that is part of an energy federation? To us, this represents a desirable future that we hope others will want to live in.

## ACKNOWLEDGMENTS

We thank Eve Schooler and everyone in Orkney who has collaborated with Watts over the years, and shared so much of their expertise. This paper is inspired by them.